\documentclass[10pt,a4paper]{article}
\usepackage{jheppub}
\usepackage{mathtools}
\usepackage{amssymb}
\usepackage{amsfonts}
\usepackage{latexsym}
\usepackage{cancel}
\usepackage{pdfpages}
\usepackage{graphicx,color}

\def\nn{\nonumber}

\def\al{\alpha}

\def\be{\beta}

\def\om{\omega}

\newcommand*{\mpl}{M_{\rm Pl}}

\newcommand{\bee}{\begin{equation}}
\newcommand{\een}{\end{equation}}

\def\bea{\begin{eqnarray}}
\def\eea{\end{eqnarray}}
\def\ba{\begin{eqnarray}}
\def\ea{\end{eqnarray}}

\def\2{\sqrt{2}}

\def\be{\begin{equation}}
\def\ee{\end{equation}}
\def\bea{\begin{eqnarray}}
\def\eea{\end{eqnarray}}

\def\om{\omega}

\title{Black Holes and  Abelian Symmetry Breaking}

\author{Javier Chagoya${}^{1}$, Gustavo Niz${}^{2}$,  Gianmassimo Tasinato${}^{1}$ }
\affiliation{${}^{(1)}$ Department of Physics, Swansea University, Swansea, SA2 8PP, U.K.}
\affiliation{${}^{(2)}$ Departamento de F\'isica, Universidad de Guanajuato - DCI, C.P. 37150, Le\'on, Guanajuato, M\'exico.}

\abstract{
Black hole configurations offer insights on the non-linear aspects of gravitational theories, and can suggest testable predictions for modifications of General Relativity. In this work, we examine exact black hole configurations in vector-tensor theories, originally proposed to explain dark energy by breaking the Abelian symmetry with a non-minimal coupling of the vector to gravity. We are able to evade the no-go theorems by Bekenstein on the existence of regular black holes in vector-tensor theories with Proca mass terms, and exhibit regular black hole solutions with a profile for the longitudinal vector polarization, characterised by an additional charge. We analytically find the most general static, spherically symmetric black hole solutions with and without a cosmological constant, and
study in some detail their features, such as how the geometry depends on the vector charges.  We also include angular momentum, and find solutions describing slowly-rotating black holes. Finally, we extend some of these solutions to higher dimensions.
}

\begin{document}
\maketitle

\flushbottom

\section{Introduction}

 The physics of black holes provides fundamental insights for understanding the nature of  gravity in  the non-linear   regime. 
    The   direct detection by Advanced LIGO of gravitational waves produced by the GW150914 event \cite{Abbott:2016blz} opens new opportunities for direct tests of black hole properties, and more 
    generally, to test the theory gravity.
      At the dawn  of this new observational era, it is  important to theoretically 
       understand   how much the physics of black holes depends
       on the theory of gravity under consideration.
%

       %

\smallskip

   While   classic results have been established in the past decades,  there are   open questions on black holes  within  Einstein gravity coupled to other fields, or in modified gravity scenarios.
  The no-hair conjecture    states that black holes are characterised by at most three quantities: mass, angular momentum, and electromagnetic charge. Including a scalar field does not normally add a supplementary conserved charge to the black hole configuration. See e.g. \cite{Herdeiro:2015waa} for a recent review on the status of the conjecture, what has been proved so far,  and possible counterexamples to it.    

If we modify our theory of gravity, Birkhoff theorem does not necessarily apply, and black hole configurations may result  different than those in GR. 
Recently, there has been an increasing interest of going beyond GR, 
 considering new scalar-tensor or vector-tensor theories of  gravity 
aimed to describe dark energy or inflation \cite{Clifton:2011jh}.  The most interesting 
 among these approaches exhibit a phenomenon called 
 screening mechanism.  Light fields can drive present day cosmic acceleration, and at the same time hide their presence thanks to screening mechanisms,
 which   suppress fifth force effects and satisfy stringent Solar system constraints. See e.g. \cite{khourylectures} for a review on this subject. 
   One of the most well studied  realizations of screening is the Vainshtein  mechanism, first proposed in \cite{vainshtein} in the context of Fierz-Pauli massive gravity: the effect of  graviton scalar polarization is concealed by its own non-linear self derivative-interactions. A broad class of theories exhibiting Vainsthein screening involve  Galileons or their extensions \cite{babichevreview}, and more generally one can construct an EFT from the Horndeski scalar-tensor set-up which captures most of the Vainshtein screening terms of other well known theories \cite{EFTVainshtein}.
 Models exhibiting (Vainshtein) screening exploit non-linear effects in the scalar sector, but have been so far mostly  analysed only in  the weak gravity regime -- where the tensor sector is treated linearly -- and around spherically symmetric sources (see \cite{davis,Chagoya:2014fza,Burrage:2014daa} for exceptions though). See \cite{Sakstein:2015aqx} for consequences of the Vainshtein mechanism in astrophysics.

The study  of  black holes in theories of modified gravity which exhibit a  screening phenomenon  is 
 interesting for various reasons. 
 On one hand, it provides
   configurations  where tensor non-linearities are important,
 hence new arenas where screening effects can be tested, possibly
  in new ways. On the other hand, 
  novel examples of black hole solutions in consistent theories where gravity is non-minimally coupled with additional fields, can shed further light on  the  generality of
  the  black hole no-hair conjecture, and suggest
  new  testable features of black hole physics.

 Important  results have been obtained so far for explicit examples of black holes in  scalar-tensor theories with screening properties, as Galileon or Horndeski systems.
In  \cite{Hui:2012qt}, the authors derive a no-go theorem against  the existence of static configurations in theories equipped by  appropriate shift symmetries. Ways out to this conclusion have been found in \cite{Rinaldi:2012vy,Sotiriou:2014pfa}, violating 
     some of the hypothesis of the no-go theorem, 
     and furnishing explicit black hole configurations.  Alternatively, simple black hole  solutions have
       been determined in  \cite{Babichev:2013cya} by considering 
        time dependent scalar configurations.  Today, 
         there are already several results on black holes within  Horndeski-like theories, including generalisations of 
         the original solutions, and studies of possible
          applications to astrophysics; see for example  \cite{Volkov:2016ehx,Herdeiro:2015waa, Herdeiro:2015gia,Sotiriou:2015pka,Silva:2016smx} for recent reviews.

\smallskip

Black hole solutions in vector-tensor theories of gravity have been less investigated recently, 
 although 
 it is an equally interesting subject with good physical motivations. When coupled with gravity,  Abelian vector fields
  provide a charge, and an  associated long range force,  to   black hole configurations.
 Here we focus on theories which break
Abelian symmetry, in which the vector longitudinal polarization becomes dynamical. 
 We investigate
 whether black hole solutions can exist in these theories. 
 %
%
%
%
%
  We have to start facing a powerful  no-go theorem: Bekenstein proved that if the
     Abelian symmetry is broken by an explicit vector mass  term, then regular  black holes do not exist   \cite{Bekenstein:1971hc}, \cite{Bekenstein:1972ny}, \cite{Bekenstein:1972ky}   (see also \cite{Adler:1978dp} for a generalization).
  We find ways out considering 
    the 
      class of vector theories breaking  Abelian symmetry, introduced in  \cite{Gripaios:2004ms,Tasinato:2014eka, Heisenberg:2014rta}, which do not need to rely on mass terms.
      The theories we consider are  related to Galileons, are free of ghostly extra modes, and  are known to exhibit screening effects in appropriate cases \cite{Tasinato:2014eka, DeFelice:2016cri}.   
       They can  be consistently covariantized  when coupled with gravity \cite{Tasinato:2014eka, Heisenberg:2014rta}, leading to  second order equations
     of motion (see also \cite{rest} for further developments on these or related theories). 
     The no-go theorems of \cite{Bekenstein:1971hc} are  circumvented by the fact that, in the cases that we consider,  Abelian symmetry is broken only by appropriate couplings with gravity.
%
   %
%
%
     We determine exact, regular black hole solutions for the specific vector-tensor set-up that we consider, with interesting properties that we summarize here:
    \begin{itemize}
    \item In absence of a bare cosmological constant, the requirement of asymptotical flatness uniquely fixes   the coupling  between vector
    and gravity. We are able to explicitly solve all the field equations, and find the most general static spherically symmetric solution for the system.  Its geometry coincides with the Schwarzschild 
    black hole, but with both transverse and longitudinal polarizations turned on for the vector.  We examine the properties of the system, and on which extent our configuration violates the no-hair conjecture. 
    \item When including a bare cosmological constant, all the field equations can again be solved exactly, and we determine the most general, static
    spherically symmetric solution. We investigate in  detail
    the  resulting geometry, which  depends both on the value of the cosmological
    constant, and ona  charge controlling the profile of the vector's longitudinal polarization.  
    \item We find and analyse exact solutions corresponding to slowly-rotating black holes, in absence of a cosmological constant. In this case, 
    the geometry corresponds to the slowly-rotating  Kerr metric, and several of the vector components have to 
    be turned on in order to solve the field equations.
       \item We extend the static, asymptotically flat solutions to higher dimensions, finding results qualitatively similar to the four dimensional case.
        \end{itemize}

\section{System under consideration}\label{sec-sys}

We  consider the following action 
\begin{equation}
S\,=\,\int d^4x \sqrt{-g}\left\{ \frac{\mpl^2}{2}\,
 R - \frac{1}{4} F^2 -\Lambda+ \beta\,\left[ (D_\mu A^\mu)^2 
                - D_\mu A_\nu D^\nu A^\mu -\frac{1}{2}A^2 R\right]  \right\}\,.\label{act1}
\end{equation}

This theory  describes Einstein gravity with a cosmological constant,  supplemented with standard, gauge invariant kinetic term for a 
vector field $A_\mu$ (whose field strength is 
$F_{\mu\nu}\,=\,\nabla_\mu A_\nu-\nabla_\nu A_\mu$), plus an additional contribution weighted by a dimensionless constant $\beta$,  breaking the  Abelian gauge symmetry by coupling the vector $A_\mu$ with gravity. 
 The first  two terms inside the square parenthesis, depending on the
 derivatives of the vector potential, are a total derivative in flat space when gravity is turned off.  To appreciate better this fact,
 after an integration by parts (and neglecting
 boundary terms), we can re-express
  the action $S$ as

 \begin{eqnarray}
 S
 &=& \int d^4x \sqrt{-g}\left[ \frac{\mpl^2}{2}\, R  -\frac{1}{4} F^2 -\Lambda + \beta\,G_{
 	\mu\nu}A^\mu A^\nu \right],\label{actG}
 \end{eqnarray}
with $G_{\mu\nu}$
 the Einstein tensor. The expression \eqref{actG}  makes  manifest  that we are breaking the gauge symmetry $A_\mu \to A_\mu+\partial_\mu \xi$  due to a direct,
 non-minimal 
  coupling of the  Maxwell potential $A_\mu$ with gravity.  The theory propagates five degrees of freedom: two helicity-2 modes (tensors) associated with the gravity sector; two helicity-1 modes  (often called vectors), associated with the transverse components  of the the vector potential  $A_\mu^T$; and one helicity-0 (scalar), the longitudinal component of the vector. 
  The dynamics of the longitudinal vector polarization is turned on by the coupling proportional to  $\beta$.
  Due to specific couplings of gravity to the vector, no additional
 sixth  mode propagates \cite{Tasinato:2014eka, Heisenberg:2014rta, Hull:2015uwa}. On the other hand, the sign of the kinetic terms of the  physical modes -- in particular the vector longitudinal polarization -- depends on the background under consideration.

 This is the simplest among the theories studied in \cite{Tasinato:2014eka, Heisenberg:2014rta} which break the vector Abelian symmetry, in such a way to avoid introducing a ghostly sixth mode. Nevertheless, it has a sufficiently
 rich structure for our purposes.  
 The field equations derived from \eqref{act1} are  (indices between brackets are symmetrized) 

\begin{eqnarray}
\frac{\mpl^2}{2} G_{\mu\nu} &=&\frac{1}{2}\left[ F_{\mu \rho} F_{\nu}{}^\rho - \frac{1}{4} g_{\mu\nu} F^2 \right]-\beta\left[  
\frac{1}{2}g_{\mu\nu}(D_\al A^\al)^2 - 2 A_{(\mu}D_{\nu)}D^\al A_\al + g_{\mu\nu} A_\al D^\al D^\beta A_\beta \right. \nn \\
& &   \left.  +\frac{1}{2} g_{\mu\nu}D_\al A_\beta D^\beta A^\al  - 2 D^\al A_{(\mu}D_{\nu)}A_\al + D_\al\left( A_{(\nu}D_{\mu)}A^\al +A_{(\mu} D^{\al} A_{\nu)}   -A^\al D_{(\mu}A_{\nu)} \right)  \right. \nn \\
& & \left. -\frac{1}{2}\left( A^2 G_{\mu\nu} + A_\mu A_\nu R  - D_\mu D_\nu A^2 + g_{\mu\nu}\square A^2   \right)\right] 
\,. \label{eeh}
\\
D^\mu F_{\mu\nu}&=& -2 \beta  G_{\mu\nu} A^\mu  \label{veh}
\,.
\end{eqnarray}

The right hand side of eq \eqref{eeh} corresponds to the energy momentum tensor (EMT) for this theory. 
The  Abelian symmetry is only broken in curved space-times. Actually, in the weak gravity regime, where $G_{\mu\nu}$ is negligible, the Abelian
 symmetry is conserved to a very good approximation. 
 Similar theories with vector fields coupled with curvature -- although not with the Einstein tensor, the only coupling that ensures the absence of a ghostly sixth mode -- have been used in cosmology 
 (see e.g. \cite{Turner:1987bw,Golovnev:2008cf}); here we investigate the existence and properties of black holes. 

Back in 1971, Bekenstein proved the non-existence of black hole solutions in theories breaking Abelian symmetry by a mass term \cite{Bekenstein:1971hc}. In few words, Bekenstein's argument starts with  noticing  that Proca's mass term  leads to a non-vanishing `right hand side' (RHS) for the vector equation (the equivalent of our equation \eqref{veh}). 
 When static configurations are considered, this
  RHS term does not allow one to turn on time-like and space-like components $A_0$ and $A_i$ simultaneously, since this would break the time-reversal symmetry which characterises the system. 
    If we analyse electrically charged configurations, with $A_0$ turned on, this implies that $A_i$ must vanish; then, a clever consideration based on the properties of a certain integral allows
      Bekenstein to conclude that regular 
black holes cannot exist in this theory, without even having  to start to solve the field equations.
  See also \cite{AyonBeato:2002wb} for additional explanations.
     
   In our work, we break the Abelian symmetry {\it not} by a mass term, but by a direct coupling with gravity, leading
   to the vector field equation  \eqref{veh}.
    This is how we avoid Bekenstein's argument, since -- as we will explicitly show -- {the longitudinal mode of the vector field
    is coupled to gravity in such a way that the relevant part of} the right hand side of equation  \eqref{veh} automatically vanishes for our static configurations. This fact implies that we can consistently turn on both the $A_0$ and $A_i$ components, without violating the time-reversal symmetry of the  system, hence  avoiding Bekenstein's arguments.
    Moreover, we find  that the additional Abelian symmetry breaking contribution in the action \eqref{actG} have important effects in determining the global properties of the spherically symmetric configuration.  The energy
   momentum tensor in \eqref{eeh}, associated with the gauge field,  backreacts on the geometry so  to modify the Reissner-Nordstr\"om charged
   black hole solution \footnote{
     As an aside comment, there is another way to avoid Bekenstein's arguments by considering asymptotically
    Lifshitz configurations. Indeed, exact black hole configurations for massive vector fields have been
    found in these theories \cite{Pang:2009pd}. Intriguingly, these solutions have some properties in common with ours, 
    as we will comment later.}.

\section{Static, spherically symmetric, charged black hole solutions}

In this section we study static,  spherically symmetric solutions for the system described by action \eqref{act1}. Our Ansatz for the metric
 and the vector field configuration is
\begin{eqnarray}
ds^2 &=& -f(r) dt^2 + h(r)^{-1} dr^2 + r^2 d\theta^2 + r^2 \sin^2\theta d \varphi^2, \\ \nonumber\\
A_\mu &=& (A_0(r), \pi(r), 0,0).
\end{eqnarray}
This is the most general static, spherically symmetric Ansatz allowing for an electrically
 charged configuration, associated with $A_0(r)$.  
Notice that we also allow for a non-trivial profile of $\pi(r)$; this is a key difference with respect to  
the usual Reissner Nordstr\"om black hole. A non-trivial profile 
for $\pi(r)$ is associated with the vector longitudinal polarization, which has an important
role in our scenario that breaks Abelian symmetry. 

Given our field Ansatz,  
the equations of motion for the system can be expressed as 
\begin{subequations}\label{eqs}
\begin{eqnarray}
0&=& \beta \pi\left[h-f (f+rf')^{-1}\right] , \label{eqs1} \\
0&=& 2 f [\mpl^2( h-1)+r^2\Lambda]+ h r^2 {A_0'}^2+ 2 \mpl^2 r h  f' \nn\\ 
&&+\beta  \left[2 A_0^2 (h-1)+2 f h (3 h-1) {\pi}^2+8 {A_0} h r {A_0}'
-2 A_0^2 h r \frac{f'}{f}+6 h^2 {\pi}^2 r f'\right]
, \label{eqs2} \\
0&=&2 f[\mpl^2(1-h-rh')-r^2\Lambda ]-h r^2 {A_0'}^2 
\nn \\ 
&&+\beta  \left[
2 A_0^2 \left(h-1+r h'\right)-2 f h {\pi} \left({\pi} \left(h+1+3 r h'\right)+4 h r {\pi'}\right)\right] ,
\label{eqs3} \\
0&=&
4 {A_0} \beta  \left(h-1+r h'\right)
-r \left[{A_0'} \left(h r \frac{f'}{f}-4 h-r h'\right)-2 h r {A_0''}\right]
. 
\label{eqs4}
\end{eqnarray}
\end{subequations}
where a prime $'$ indicates derivative along the radial direction.  
When $\beta\neq 0$ and the scalar profile is non-vanishing,
the first equation determines $h$ only in
terms of $f$. As a consequence, the Ricci scalar can be generally expressed as
\begin{equation}
R=\frac{\left(2 f-r f'\right) \left(2 f'+r f''\right)}{2 r \left(f+r f'\right)^2}.
\end{equation}
Therefore, if $f$ asymptotically scales with some power of $r$, $f\sim r^n$ for large $r$ (with $n\neq-1$) then 
 the Ricci scalar  vanishes  at infinity, $R\sim 1/r^{2}$. 

We start discussing solutions in absence of cosmological constant. 


\subsection{Asymptotically flat configurations in absence of a cosmological constant }\label{sec-flat}


\smallskip
 When $\beta=0$, standard steps lead to Reissner-Nordstr\"om configuration as a unique electrically charged, spherically symmetric solution for the field equations
\bea
f&=&h\,=\,1-\frac{2 M}{r}+\frac{Q^2}{2 \mpl^2 r^2},
\\
A_0&=&\frac{Q}{r}+P,
\\
\pi&=&0.
\eea
This configuration is asymptotically flat. Since gauge symmetry is preserved, the vector longitudinal mode profile $\pi(r)$ is not physical and can be switched off.  
In the previous formulae, $M$ is the black hole mass and  $Q$ is the black hole electric charge.
 Notice that the additional integration constant $P$ in the vector profile has no physical effects, and can be set
 to zero thanks to gauge symmetry. 
 On the other hand, the electric charge $Q$
   affects the metric introducing a $Q/r^2$
correction to the Schwarzschild metric; the geometry `feels' the electric charge. See \cite{Carroll:2004st} for a textbook
 discussion, including an extension to magnetically charged configurations.

\smallskip

However, when $\beta\neq0$, the structure of the solution is very different. 
 We require   that the metric  asymptotically approaches
 Minkowski space. This means that we focus on the large $r$ limit of
 the configuration,  where we can approximate $f\,=\,h\,=\,1$. Substituting  $f\,=\,h\,=\,1$ in equations \eqref{eqs1}-\eqref{eqs4}, we find that they can be satisfied for a non-trivial
 profile for $A_0$, but {\it only if}\,
 \be
 \beta=\frac14\,. \label{betach}
 \ee
 
 So the requirement of approaching a flat space asympotically, singles out a unique value for the coupling constant~\footnote{We also
  checked that requiring  more generic asymptotic behaviours for $f$ and $h$ -- for example demanding a power-law -- also fixes (different) values for $\beta$. However, it is then not possible to find 
  simple solutions for the complete equations for all values of $r$.}.
   Remarkably, after making the choice \eqref{betach},
 we can return back to the set of equations \eqref{eqs1}-\eqref{eqs4} and find  that
  a simple manipulation of them~\footnote{We can solve eq \eqref{eqs1} for $h$, and substitute the results in the remaining equations.  \eqref{eqs3} provides a simple differential equation for $A_0$, with \eqref{g1a0}  as unique solution. Substuting this profile in eqs  \eqref{eqs2},  \eqref{eqs4}, we find simple linear differential equations leading to \eqref{g1f}, \eqref{g1pi}
  as general solutions.  There is no solution for $\pi(r)=0$. } provides a {\it unique} static solution, which turns out to be asymptotically flat. The 
   most general 
  solution reads
 \bea
f&=&h\,=\,1-\frac{2 M}{r}  \label{g1f}
\\
A_0&=&\frac{Q}{r}+P \label{g1a0}
\\
\pi&=&\frac{\sqrt{Q^2+2\,P\,Q\,r+2 \,M\,P^2\,r}}{ r-2 M }  \label{g1pi}
\eea
where $M$ is the black hole mass, while $Q$ and $P$ indicate charge parameters. 
This unique static solution has various interesting properties, which can be summarised as follows: 

\begin{itemize}
\item[$\blacktriangleright$]  The geometry is described  by the 
 Schwarzschild solution characterised by $M$, and it is independent of  the charges $Q$ and $P$. The Abelian symmetry breaking  self-interactions contribute
 to the  energy momentum tensor in such a way to {\it exactly} compensate  for the contributions of the vector charges to the geometry. In some sense, 
 the vector self-interactions are able to screen the vector  charges $Q$ and $P$ from the geometry, which is insensitive to the presence of the vector. Such a
 behaviour was also found in scalar-tensor black holes \cite{Babichev:2013cya} for
 a similar system, leading to so-called stealthy Schwarzschild configurations.
    Also, in asymptotically Lifshitz
 systems \cite{Pang:2009pd} with massive vectors the geometry does not depend on the vector profile.

\item[$\blacktriangleright$] 
Since the geometry  corresponds to a Schwarzschild configuration, the corresponding
 Einstein tensor vanishes, implying that the right hand side of the vector field equation \eqref{veh} is zero.
 As we explained in Section \ref{sec-sys}, 
 this avoids  Bekenstein's arguments \cite{Bekenstein:1971hc}, since we can have both $A_0$ and $\pi$ switched on without breaking 
 the time reversal symmetry, allowing us to find a regular black hole
 configuration in an Abelian symmetry breaking theory.

\item[$\blacktriangleright$]  
Each of the curvature invariants, and the components of the EMT, are well behaved for $r>0$. No 
essential singularities are induced on the system by the new vector interactions, besides the singularity at the origin,
 covered  by the Schwarzschild horizon. The  singularity at $r=2M$ in the scalar profile $\pi(r)$ is  only apparent -- analogous
to the apparent singularity in the horizon of the geometry -- and does not appear when computing the components of the EMT. 

\item[$\blacktriangleright$]  The vector profile $A_0(r)$ and the scalar profile $\pi(r)$ depend on two independent integration constants, $Q$ and $P$.
The profile for  $A_0(r)$  leads to a long range electric field sourced by the black hole, identical to the
 Reissner-Nordstr\"om configuration. 
 While,  as we have seen, the parameter  
 $P$ has no physical implications for  the
Reissner-Nordstr\"om black hole, 
 in this case  it is a charge controlling  the scalar profile $\pi(r)$ (although not the geometry). 
 Depending on the presence or not of $P$, the scalar profile $\pi(r)$ scales as $1/\sqrt{r}$ or $1/r$
 at large distances. 
   Notice also that we can turn off the charge $Q$ -- so to switch off the long range electric field -- and leave the
  charge $P$, characterising a non-trivial
   scalar profile,  turned on  in the presence of a  black hole mass.

  In this sense, $P$ represents an additional charge for the configuration,
  besides mass and electric charge. However this charge is not `seen' by applying the standard Gauss law -- since
  the field $\pi$ does not contribute to $F_{\mu\nu}$ -- hence strictly speaking
  this configuration  does not violate the no-hair conjecture.  
  On the other hand,   the scalar
 profile might be probed  by some test object which couples to the vector field $ A_\mu$ in a way which does
 not respect the
 Abelian symmetry.
 It would be interesting to analyse specific systems
  which can probe in this way the longitudinal component of $\pi$. 
  This is an interesting, but more model dependent issue that we leave for future work.
  
\end{itemize}
We explicitly checked that our configuration is stable under  perturbations that are spherically symmetric (but time dependent), provided that
$Q$ and $P$ are small.  
In the next subsection, we discuss how the addition of a cosmological
constant drastically changes the properties of the geometry.

\subsection{Configurations including a cosmological constant}
\label{sec-cc}

We now consider how including  a cosmological constant affects
our configuration. Interestingly,  the  equations of 
motion determining a static, spherically 
symmetric configuration,  can again be solved straightforwardly in the case of 
  $\beta =1/4$. Making this choice for this parameter, with a procedure  analogous to 
  the one of Section \ref{sec-flat}, we find that the most general solution of our set of equations is
\begin{subequations}
	\begin{eqnarray}
	f &=& 1 -\frac{2M}{r} 
	+\frac{4 r^2 \Lambda _P}{3}+\frac{4}{5} r^4 \Lambda _P^2 \, , \label{solc-f}\\
	h &=& f\,\left(1+2 r^2 \Lambda _P\right)^{-2} ,\label{solc-h}\\
		A_0 &=& \frac{Q}{r}+P\left( 1+\frac{2}{3} r^2 \Lambda_P  \right),\label{solc-a0}\\
	\pi&=& 
	\sqrt{\frac{A_0^2}{f h }-
	\frac{P^2(1+2 r^2 \Lambda_P)-8 \mpl^2 r^2 \Lambda_P }{h}}\label{solc-pi} 
\\
&= & \frac{1+2 r^2 \Lambda_P}{\sqrt{f}}\left[\frac{\left(3 Q+r P  \left(3+2 r^2 \Lambda _P\right)\right){}^2}{9 r^2 f}-P^2(1+2 r^2 \Lambda_P)+8 \mpl^2 r^2 \Lambda _P   \right]^{1/2},
	\end{eqnarray}
\end{subequations}
where 
\be
\Lambda_P \equiv\frac{ \Lambda}{P^2-4 \mpl^2}\,, \label{lap}
\ee
 and $Q$ and $P$ are integration constants (with $P^2\neq4 \mpl^2$ to avoid a singular geometry).  
 The presence of the cosmological constant drastically changes the profile of the solution 
 we analysed in section \ref{sec-flat}. The properties of the new configuration are the following:
 \begin{itemize}
 \item[$\blacktriangleright$] The geometry now depends on the scalar charge $P$, and
 the cosmological constant, through the parameter $\Lambda_P$
  introduced in \eqref{lap}. The metric functions $f$ and $h$ differ one from the other, and the geometry appears different
  from (a)dS space, since the metric component $f$ scales as $r^4$ at large values of $r$. 
  The profiles for $A_0$ and $\pi$ do not generally vanish for large values of $r$. Notice that if $Q=P=0$ the
  electric field controlled by $A_0$ is turned off, but the longitudinal profile  is still non-trivial, thanks to the contribution
  of the cosmological constant.
  
  \item[$\blacktriangleright$] 
  The Ricci scalar reads
  \begin{equation}
R=-\frac{8 \Lambda _P (15 M-5 r+4 \Lambda _P^2 r^5 )}{5 r (1+2 \Lambda _P r^2 ){}^3}\,.
\end{equation}
 If $\Lambda_P>0$, the only essential singularity is located at the origin $r=0$. If $\Lambda_P<0$, there is an additional essential singularity
 when 
 \be
 1+2 r^2 \Lambda_P =0\,.
\ee
As we will discuss in the next point, such singularity can be covered by an horizon, for appropriate choices of the parameters.
 The Ricci scalar vanishes at infinity; but this is not the case for all the components of the Riemann and Ricci tensor. In particular the Einstein tensor scales as $r^2$ at large distances, similarly to maximally symmetric space-times. It would be interesting to study how all these 
non-vanishing components of the curvature affect the propagation of matter and waves through space-time in our set-up.

We checked that the combination $G_{r r} \,A^r$ vanishes for our solution. This implies that the right hand side of the $r$ component of eq \eqref{veh} vanishes:
as explained in Section \ref{sec-sys}, this avoids Bekenstein's no-go arguments against the existence of black hole solutions.

 \item[$\blacktriangleright$] It is possible to prove that the geometry is characterised by a unique horizon, corresponding to the unique real zero of the function $f$
 in \eqref{solc-f}. Indeed, the quantity $r f(r)$ is a monotonic function of $r$, which is negative for $r$ very small (assuming $M>0$) and positive for $r$ very large (this is
 immediate to see if $\Lambda_P>0$, and simple to prove if $\Lambda_P<0$). Hence
 it must vanish once for some intermediate value of $r$, corresponding to the black hole horizon.
 As we have seen earlier, if $\Lambda_P\geq 0$ the configuration is free
 from essential singularities for $r\neq 0$.  If $\Lambda_P< 0$ instead we have an essential singularity at the radius
$$ r_S=\frac{1}{\sqrt{-2\Lambda_P}}. $$
To determine whether this singularity is covered by the black hole horizon or not, we evaluate the radial metric function $f$ at
$r_S$. We find  that the condition for ensuring that the  singularity is inside the horizon is $\sqrt{-2\Lambda_P} M > 4/15$.

  \item[$\blacktriangleright$] 
A  change of variable 
    \be
   \rho\equiv \left(r+\frac23 r^3 \Lambda _P\right) \label{rhor}
   \ee
   allows us to express the metric in the following nicer  form:
   \be
   d s^2\,=\,-f\left[ r(\rho) \right]\,d t^2+\frac{d \rho^2}{f\left[ r(\rho) \right]}+r^2(\rho)\,d \Omega^2
   \ee
   However, since relation \eqref{rhor} can not be easily inverted, the resulting expression for 
   $f\left[ r(\rho) \right]$ is not particularly illuminating.

    \end{itemize}

\noindent
This concludes our discussion of the spherically symmetric configurations
in our theory. 

\section{Slowly rotating black hole configurations}

An open issue of screening mechanisms is to understand what happens when one renounces to 
spherical symmetry (see \cite{davis,Chagoya:2014fza} for some insights). Hence it is interesting to investigate  the behaviour  
of  stationary black hole configurations that are only axially symmetric, and not spherically symmetric.
 A physically relevant case are rotating black holes. 

 In our set-up, when turning off the cosmological constant, we  are able to determine a solution for the field   equations only for  the case of slowly rotating configurations. 
A slowly rotating spacetime can be described by the metric
\begin{eqnarray}\label{rotmet}
ds^2 &=& -f(r) dt^2 + h(r)^{-1} dr^2 + r^2 d\theta^2 + r^2 \sin^2\theta d \varphi^2 + 2 a r^2 \om(r) \sin^2\theta\, d t \,d\varphi,
\end{eqnarray}
where $a$ is a small parameter,  which controls the amount of rotation. We are
interested in configurations that depend at most linearly on the small parameter $a$.

We are not able to find the most general solution in this case, hence we proceed as follows. Building on the results
of the previous sections, in the small $a$ limit we make the hypothesis that the geometry coincides with the 
small rotation limit of Kerr black holes -- that is, there is not backreaction of the vector charge to  the geometry. 
This expectation turns out to be correct, and the field equations admit a solution for the geometry corresponding
to slowly rotating Kerr: $f$, $h$ correspond to the Schwarzschild geometry, while   $\omega(r)$ acquires the profile
$\om = {J}/{r^3}$, with $J$ an integration constant parameterising angular momentum. 
Several components of the vector require  non-trivial profiles for solving the equations.  The components $A_0$ and $\pi$ for the time-like 
component of the gauge potential and longitudinal scalar $\pi$ are given by the same solutions of  \eqref{g1a0}, \eqref{g1pi} with $P=0$; the profile required for 
$A_\phi$ is
\begin{equation}
 A_\phi = a\frac{J Q}{2 M r}\,.
\end{equation}
This  choice of vector components solves all the equations. 
%
%
%
%
%
These findings show that the geometry satisfying the field equations corresponds exactly to the slow-rotation limit of Kerr space-time (that
by itself satisfies Einstein equation in the vacuum), while the time and angular components of the vector field 
coincide with the slow-rotation limit of the Kerr-Newman 
solution. In addition, the Abelian symmetry breaking terms switch a non-trivial profile on for the longitudinal component of the vector.   
 Notice that  the corresponding Einstein tensor
 vanishes, and as in the previous sections we can avoid Bekenstein no-go theorem.



\section{Higher dimensional solutions}

The system we analyzed in four dimensions can straightforwardly be analysed in higher dimensions, leading again 
to an analytically manageable system of equations.
  We consider action  the \eqref{act1} 
in dimensions $d=5,\,\dots, 10$, with the spherically symmetric Ansatz
 \bea
 d s^2&=&-f(r) \,d t^2+h(r)^{-1}\,d r^2+r^2\,d  \Omega_{(d-2)}^2\,,
 \\
 A_\mu&=&\left(A_0(r),\,\pi(r),\,0,\,\dots\,, 0\right)\,.
 \eea
Also in different dimensions, asymptotical flatness requires a specific value 
for the coupling constant $\beta$, 
 which we need to be
\be
\beta\,=\,\frac{d-3}{2 \,d-4}\, .
\ee
Hence, 
the  field solutions can  be written compactly as a function of the spacetime dimension $d$ and  three integration constants $M_d$, $Q_d$, $P_d$ as
\bea
f&=&{h}\,=\,1-\frac{2 M_{d}}{r^{d-3}}\, ,
\\
A_0&=& \frac{Q_d}{r^{d-3}}+P_d\, ,
\\
\pi&=&\frac{\sqrt{Q_d^2+
2 \,M_d\,P_d^2\,r^{d-3}+ 2\,P_d\,Q_d\,r^{d-3}
}}{r^{d-3}-2 M_d}\, .
\eea
The physical properties
of these configurations are similar to those in four dimensions, thus we conjecture that these solution should exists in dimensions $d>10$, with the same formulas given above.
\section{Discussion}

We  examined  exact black hole solutions for a particular vector-tensor
theory of gravity, with the Abelian symmetry broken by a certain  non-minimal coupling 
 of the vector $A^\mu$ to gravity, given by
$$\beta \,G_{\mu\nu} \,A^\mu\,A^\nu,$$ where $G_{\mu\nu}$ is the Einstein tensor.
Our solutions are  characterised by a long range electric field associated with an electric charge $Q$, and
 by a non-trivial  profile for the vector's longitudinal polarisation, associated with a charge $P$ which however does
not contribute to a Gauss law. We construct these solutions in four and higher dimensions, 
and find that asymptotical flatness, in the absence of a bare cosmological constant, singles out a particular value for above's coupling $\beta$, given by $\beta={(d-3)}/{(2 \,d-4})$, with $d$ the number
of dimensions. 
By choosing
this value, we have been able to integrate explicitly  all the equations of motion, and found the most general solution, that can be simply extended to include a bare cosmological constant in 4d. Moreover, we determined exact solutions of the field equations for slowly rotating black holes. 
  The final expressions for the field configurations are relatively simple,
and we discussed their physical properties and implications.  We also discussed in some detail how our system is able to avoid Bekenstein's no-go 
arguments \cite{Bekenstein:1971hc} on black holes with massive vector hairs.

It would be important to check in detail whether our configurations are stable under small disturbances, possibly implementing the techniques that
have been recently developed for studying similar scalar-tensor black hole configurations \cite{Ogawa:2015pea}. We explicitly checked that our solutions
are stable under spherically symmetric fluctuations in certain cases, but a more general analysis would be needed to settle the question. 

It would also  be interesting to explore solutions which are not asymptotically flat, for which $\beta\neq (d-3)/{(2 \,d-4})$, or to consider more general theories that break the Abelian symmetry without intoducing a ghost, as the ones considered here. 
Actually, we have looked into some of these more general theories, but have not been able to find analytical solutions. However a numerical analysis of these systems might be feasible.
 
Finally,
it would also be very interesting to investigate whether these configurations can have some astrophysical relevance.  
 It is important  to study realistic black hole formation in this scenarios, and see whether they can lead
to configurations equipped by a long distance scalar hair associated with the vector longitudinal polarization. Moreover, our Abelian symmetry
breaking coupling of vector to gravity is extremely suppressed for weakly gravitating systems, hence such terms might realistically describe a coupling
between electromagnetism and gravity, with potentially distinctive effects in the strong gravity regimes involving black holes and cosmology.

\acknowledgments

It is a pleasure to thank Ivonne Zavala for discussions. GT  is partially supported by STFC grant ST/N001435/1. GN and, partially, JC are supported by the grant CONACYT/179208. JC is also supported by the grant CONACYT/263819.

\end{document}